# ADVANCED SINGLE IMAGE RESOLUTION UPSURGING USING A GENERATIVE ADVERSARIAL NETWORK

Md. Moshiur Rahman[1], Samrat Kumar Dey[2]and Kabid Hassan Shibly[3]

[1]School of Science and Technology, Bangladesh Open University, Gazipur-1705, Bangladesh
[2,3]Department of Computer Science and Engineering, Dhaka International University,Dhaka-1205, Bangladesh

***ABSTRACT***

*The resolution of an image is a very important criterion for evaluating the quality of the image. Higher resolution of image is always preferable as images of lower resolution are unsuitable due to fuzzy quality. Higher resolution of image is important for various fields such as medical imaging; astronomy works and so on as images of lower resolution becomes unclear and indistinct when their sizes are enlarged. In recent times, various research works are performed to generate higher resolution of an image from its lower resolution. In this paper, we have proposed a technique of generating higher resolution images form lower resolution using Residual in Residual Dense Block network architecture with a deep network. We have also compared our method with other methods to prove that our method provides better visual quality images.*

## 1. INTRODUCTION

Bringing out an accurate high-resolution image from a low-resolution image is very effective and renowned work in image Super Resolution research works. It has got appreciable attention in Computer Vision community and it has a huge application scope. This particular problem is acclaimed for high up scaling work where the texture detail is barely up to the mark in reconstructed images. The algorithms related to up-scaling images mainly try to minimize the cumulative squared error between the compressed and the original image. It is also called Mean Square Error (MSE). But this technique increases the performance time and maximizes peak signal- to-noise ratio (PSNR). It also cannot provide a near ground- truth similar image.

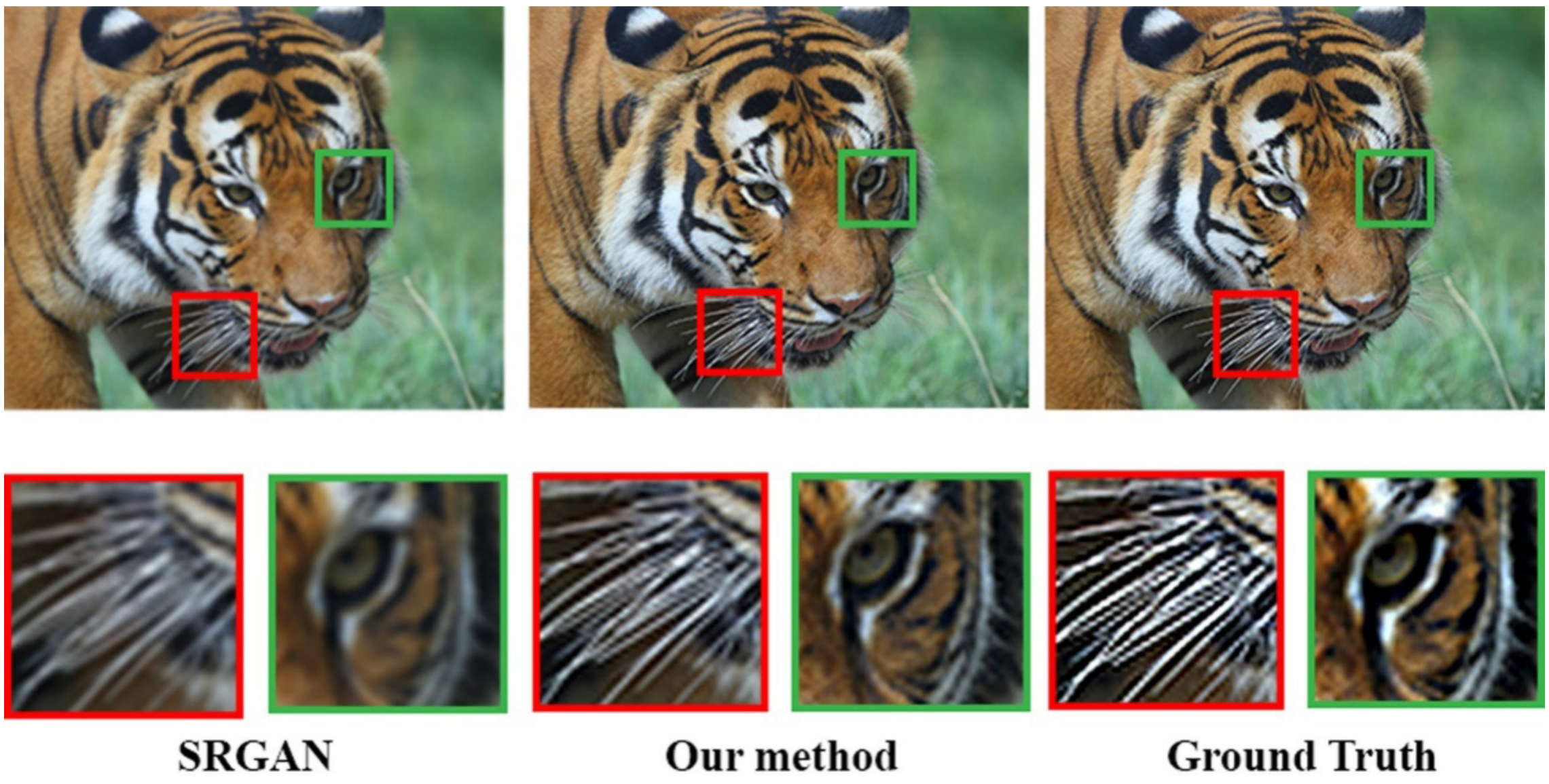

Figure 1. The super-resolution results of SRGAN, the proposed method and the ground-truth. Proposed method outperforms SRGAN in sharpness and details.

There still exits a clear gap between SRGAN and the ground-truth (GT) images as shown Figure 1. We revisit the exiting model and improve it in different aspects. Our experiment shows that this improvement helps to recreate more realistic texture details than SRGAN. Here we have employed a deeper network with RRDB network architecture with skip-connection and diverge from MSE as the sole optimization target. We have optimized the perceptual loss in GAN framework.

## 2. Related Work

### 2.1. Image Super-Resolution

Some of the overview articles on super-resolution including Yang et al, [1], Nasrollahi and Moeslund [2]. They worked on single image resolution enhancement method that we aimed in this paper. A well-known prediction-based method is bicubic or Lanczos [3]. This is very fast and provides overly smooth texture. Edge-prevention methods also have been proposed in recent time [1, 4]. Tai et al [6] and Zhang et al [7] reconstructed realistic texture detail and multi-scale dictionary. The work by Bruna et al [8] and Johnson et al [9] are relevant to our paper.

### 2.2. CNN Design & Loss Function

Deeper network can be hard to train but it increases network accuracy and better images [10, 11]. In terms of speed and accuracy and learning [12, 13, 14] papers are good. In Mathieu et al. [15] and Denton et al. [16], the authors tackled this problem by employing generative adversarial networks (GANs) [17] for the application of image generation. The idea of using GANs to learn a mapping from one manifold to another is described by Li and Wand [18] for style transfer and Yeh et al. [19] for inpainting deep learning. Some different methods have been proposed to stable train a very deep model. So, the residual path is developed to stabilize the training and improve the performance [20, 21, 22]. Szegedy et al. [23] first employed Residual scaling and also used in Enhanced Deep Residual Networks (EDSR). He et al. [24] propose a robust initialization method for general deep networks. And it is VGG-style networks without Batch Normalization (BN). To facilitate training a deeper network, we use a compact and effective residual-in-residual dense block, which also helps to improve the perceptual quality.

Photo-realism is usually attained by adversarial training with GAN. There are several recent works that focus on developing more effective GAN frameworks. To minimize a reasonable and efficient approximation of Wasserstein distance and regularizes discriminator by weight clipping proposed in WGAN [25]. Some improved regularization includes gradient clipping [26] and spectral normalization [27]. Relativistic discriminator [28] is developed to simultaneously decrease the probability that real data are real and also to increase the probability that generated data are real. In this work, we improve and enhance SRGAN by employing a relativistic average GAN.

## 3. METHODOLOGY

Our main goal is to train a generative adversarial network that estimates the corresponding High-Resolution counterpart of a given Low-Resolution image and to make sure this, we train our GAN as feed-forward CNN. Now in this section, we describe our network and then discuss the discriminator and perceptual loss. We mainly focused on two aspects, such that, i) the absent of BN layers, that means, we are not going to use Batch-Norm layers as SRGAN [29] does, ii) using RRDB (Residual-in-Residual Dense Block) from ESRGAN [30] which is depicted in Figure 2.

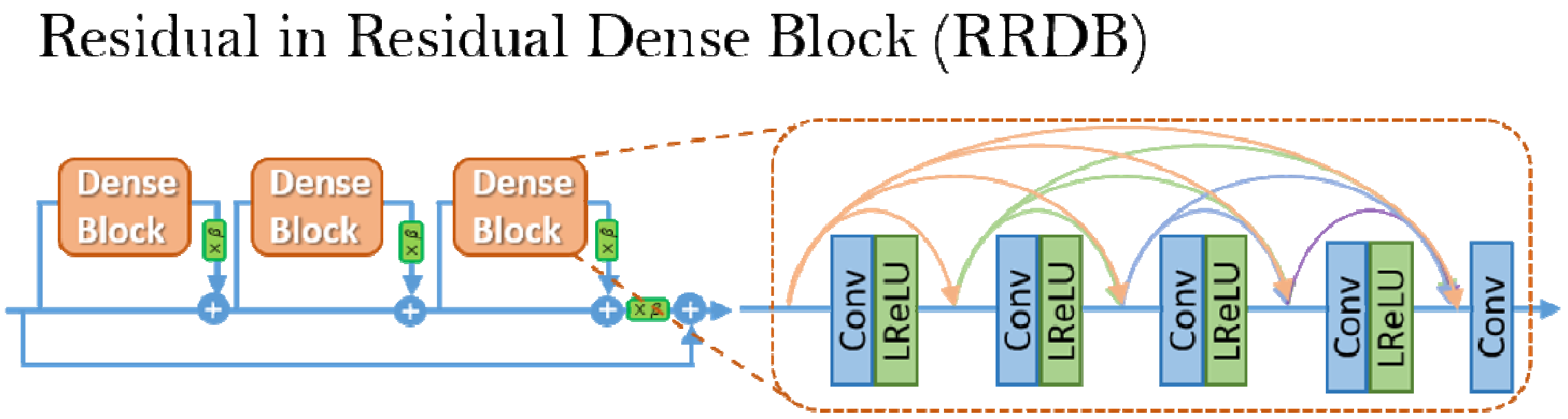

Figure 2. Our used network architecture (*RRDB*)

Here we are using GAN or Generative Adversarial Network. The adversarial model displaying structure is most direct to apply when the models are both multi-layer perceptions. Generative Networks is an approach to generative modeling using deep learning methods. Generative modeling is an unsupervised learning task in machine learning that involves automatically discovering and learning the regularities or patterns in input data in such a way that the model can be used to generate or output new examples that plausibly could have been drawn from the original dataset. However, BN layers normalize the features in a batch during training and whole dataset during testing. If the trained test dataset very different, BN layers limit the generation ability. So, we make the network deeper and trained under the GAN framework. To make a better performance and get stability overtraining we remove BN layers. It also reduces computational complexity and memory usage. To keep a high-level architecture design, we used the RRDB basic block. The study and observation say more layers and connections always improve performance. The architecture we are using has a deeper and complex architecture. So, the network capacity becomes higher than before. So, for an improved network, we are scaling down residuals by multiplying a constant between 0 and 1 and, as we empirically find residual architecture is easier to train when the initial parameter variance becomes smaller.

We also focused on enhancing discriminatory based on GAN. We are using Relativistic GAN, that can estimate the probability that one input image is real. A relativistic discriminator tries to predict the probability that a real image is relatively more realistic than a fake one. So, we used relativistic discriminator instead of a standard one. The discriminator loss function is defined in Equation (1).

$$L_D^{R_a} = -E_{x_r}[\log(D_{R_a}(x_r, x_f))] - E_{x_f}[\log(1 - D_{R_a}(x_r, x_f))] \quad (1)$$

And the adversarial loss for the generator is in Equation (2)

$$L_G^{R_a} = -E_{x_r}[\log(D_{R_a}(x_r, x_f))] - E_{x_f}[\log(1 - D_{R_a}(x_r, x_f))] \quad (2)$$

The generator benefits from the gradients from both generated data and real data in adversarial training and this discriminator helps to learn sharper edges and more detailed textures. Perceptual loss in the activation layers of a pre-trained deep network, the distance between two activated features is minimized. We can use features before the activation layers. This resolves few complications such as, activated features that are very scattered which leads to weak supervision. That results in a menial performance. Besides using the feature after activating leads to inconsistency in the reconstructed image compared to the ground-truth image. However, we used 19 layers of VGG layers which is a pre-trained network. The total loss for the generator is shown in Equation (3)

$$L_G = L_{percep} + \lambda L_G^{R_a} + \eta L_1 \quad (3)$$

## 4. EXPERIMENTAL DESIGN AND RESULT

We trained our models in RGB channels. Also, trained with the training dataset with random horizontal flips. We trained a model with the initial loss. Then we employed the trained model as an initializer for the generator. Here we use loss function in (3). Pre-training with loss based on pixel helps GAN-based methods to gain visually better results. As we are comparing with SRGAN [29], our evaluations are performed under benchmark datasets Set5 [31], Set14 [32], and BDS300 [33]. For training data, we mainly used DIV2K [34] dataset. It contains 800 images of 2K resolution in the train set. We also used Out Door Scene Training (OST) [35] dataset. The following produces sharper, more natural photo-realistic images which is shown in Figure 3 based on the method we proposed in this exploration.

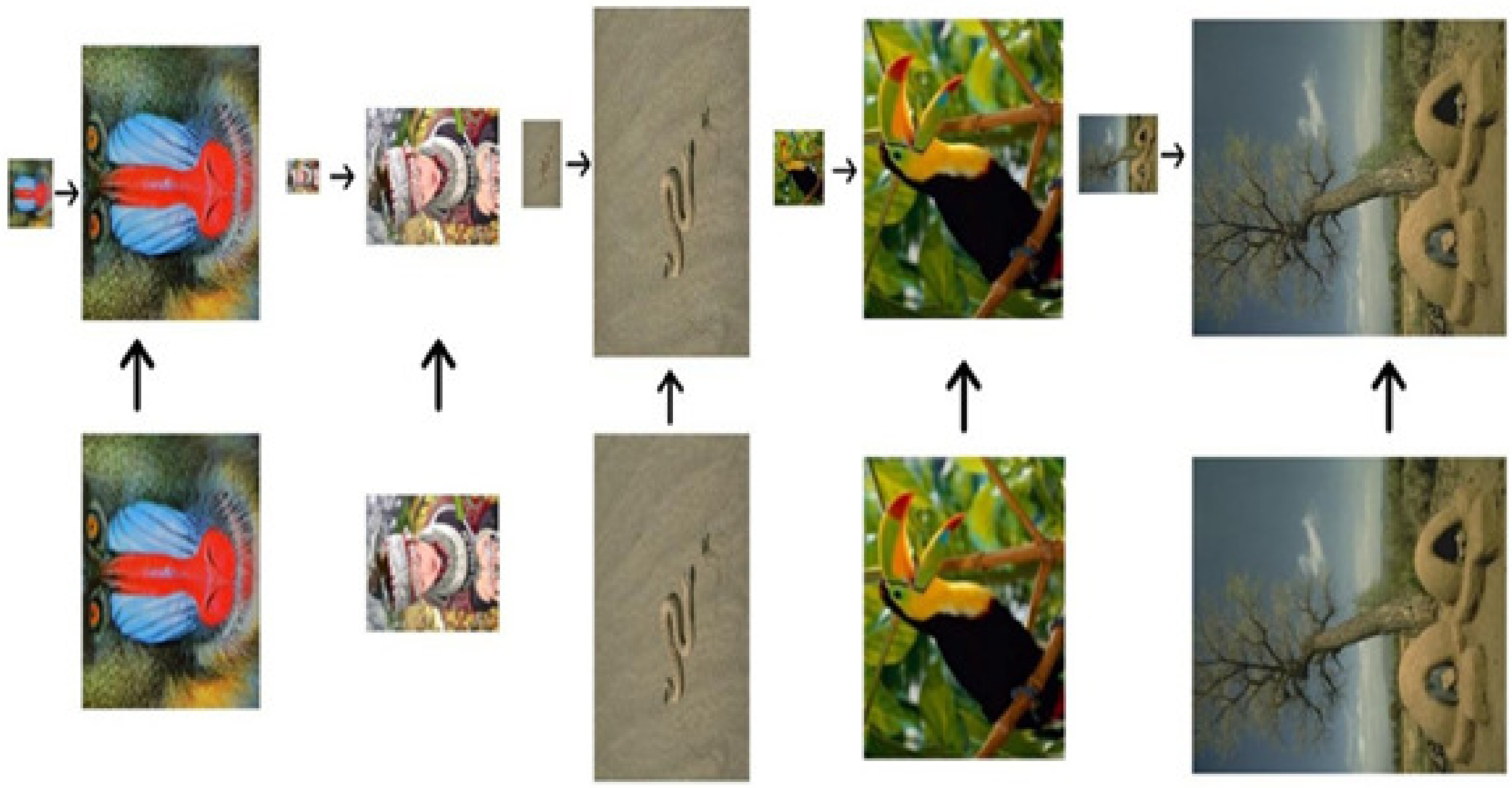

Figure 3. Low Resolution to High Resolution image using our proposed method

It can be observed the performance of our method compared to SRGAN and others which is illustrates in Figure 4. We can observe both the sharpness and detailed of each individual image. Our Proposed approach is capable of generating more detail structures.

z

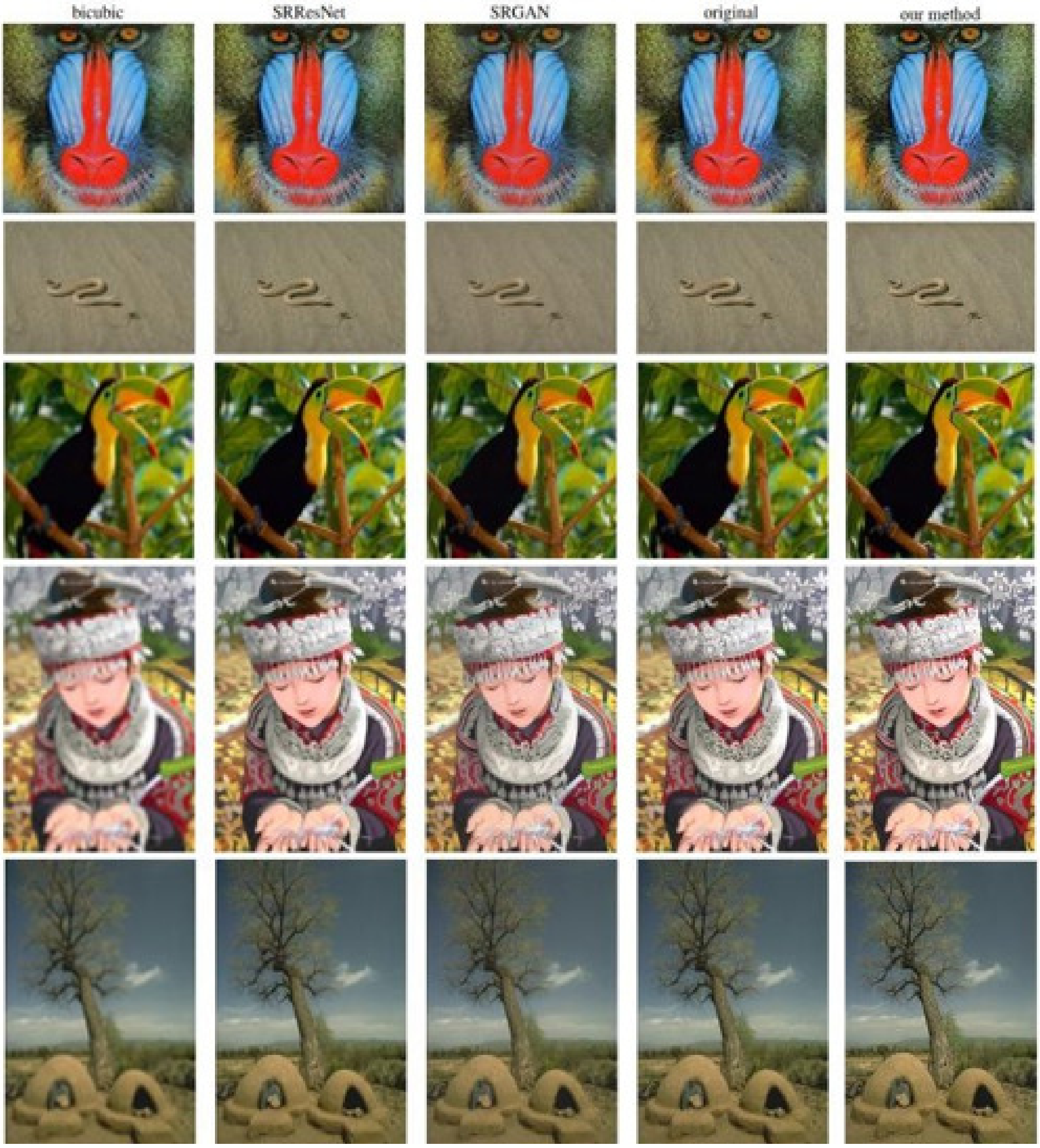

Figure 4. Comparative representation of proposed methods with existing methods

In TABLE 1. we can see the content loss in different testing standards. Such as MOS (Mean Opinion Score), SSIM (Standard Similarity Index Metrics) etc. based on different benchmark datasets.

Table 1. Loss Function Performance

| **Set5 Dataset** | **SRResNet** | |
|---|---|---|
| | **MSE** | **VGG22** |
| PSNR | 32.05 | 30.51 |
| SSIM | 0.9019 | 0.8803 |
| MOS | 3.37 | 3.46 |
| **Set5 Dataset** | **SRGAN** | |
| | **MSE** | **VGG22** |
| PSNR | 30.64 | 29.84 |
| SSIM | 0.8701 | 0.84.68 |
| MOS | 3.77 | 3.478 |
| **Set5 Dataset** | **Our method** | |
| | **MSE** | **VGG22** |
| PSNR | 29.56 | 28.12 |
| SSIM | 0.9109 | 0.8881 |
| MOS | 3.64 | 3.49 |

## 5. CONCLUSIONS

We have presented an approach that performs consistently better in quality than previous methods of Super Resolution. We have formulated a different approach using RRDB without BN layers. We also use the relativistic GAN as the discriminator. Which can guide the generator to recover more detailed texture. However, we have also enhanced the perceptual loss which offer better supervision and restore more realistic texture. In near future we will try to refine and develop the reconstructed images in a better way. We have discussed about some methods that work with open accessible datasets that are widely used. Some of the limitations with pervious methods are focused on image upscaling. We worked with some arguments the content loss function with adversarial loss training by Generative Adversarial Networks. We ensured the four times up-scaling in the reconstructed images as the SRGAN. However, in near future we will try to refine and develop the reconstructed images in a better way.

## AUTHORS

**Md. Moshiur Rahman** has been lecturing in CSE since 2017. He has received M. Sc Engineering degree in Computer Science and Engineering Islamic University of Technology (IUT), Bangladesh in 2015. His research interests include data and image mining, Semantic Web, Artificial and Business Intelligence, Educational Technology, E-learning etc. He is now serving only ODL based public University in Bangladesh named Bangladesh Open University (BOU).

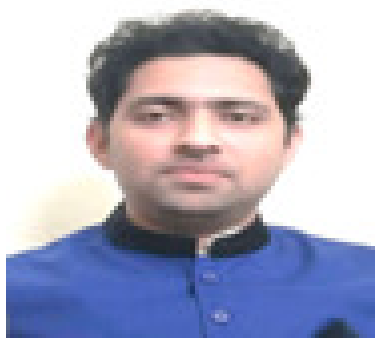

**Samrat Kumar Dey** received his Master's degree in Computer Science and Engineering (CSE) from Military Institute of Science and Technology (MIST) under Bangladesh University of Professionals (BUP), Dhaka-1216, Bangladesh, in 2019. He is an Assistant Professor with the Department of Computer Science and Engineering, Dhaka International University (DIU), Dhaka-1205, Bangladesh. His research interest mainly focuses on Visual Data Analytics, Data Science, Human-Computer Interaction (HCI), Usability and UX analysis, Machine Learning/Deep Learning, ICT4D, e-learning and Network Security.

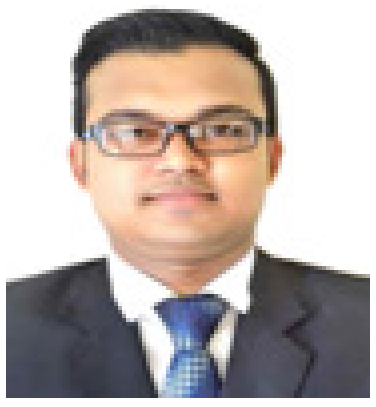

**Kabid Hassan Shibly** was an undergrad student within the Computer Science & Engineering program at the Dhaka International University. He completed his undergraduate degree in 2018. His research concentrates on computer vision, image processing and artificial intelligence.

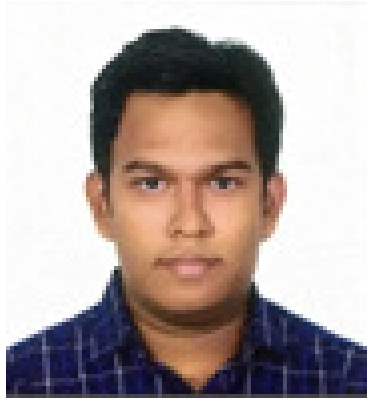